\documentclass[12pt]{article}
\usepackage{epsfig}
\usepackage{graphicx}
\usepackage{amsmath}
\usepackage{hhline}
\usepackage{amssymb}
\usepackage{times}
\usepackage{cite}

\newlength{\dinwidth}
\newlength{\dinmargin}
\begin{document}  
\newcommand{\pom}{{I\!\!P}}
\newcommand{\reg}{{I\!\!R}}
\newcommand{\slowpi}{\pi_{\mathit{slow}}}
\newcommand{\fiidiii}{F_2^{D(3)}}
\newcommand{\fiidiiiarg}{\fiidiii\,(\beta,\,Q^2,\,x)}
\newcommand{\n}{1.19\pm 0.06 (stat.) \pm0.07 (syst.)}
\newcommand{\nz}{1.30\pm 0.08 (stat.)^{+0.08}_{-0.14} (syst.)}
\newcommand{\fiidiiiful}{F_2^{D(4)}\,(\beta,\,Q^2,\,x,\,t)}
\newcommand{\fiipom}{\tilde F_2^D}
\newcommand{\ALPHA}{1.10\pm0.03 (stat.) \pm0.04 (syst.)}
\newcommand{\ALPHAZ}{1.15\pm0.04 (stat.)^{+0.04}_{-0.07} (syst.)}
\newcommand{\fiipomarg}{\fiipom\,(\beta,\,Q^2)}
\newcommand{\pomflux}{f_{\pom / p}}
\newcommand{\nxpom}{1.19\pm 0.06 (stat.) \pm0.07 (syst.)}
\newcommand {\gapprox}
   {\raisebox{-0.7ex}{$\stackrel {\textstyle>}{\sim}$}}
\newcommand {\lapprox}
   {\raisebox{-0.7ex}{$\stackrel {\textstyle<}{\sim}$}}
\def\gsim{\,\lower.25ex\hbox{$\scriptstyle\sim$}\kern-1.30ex%
\raise 0.55ex\hbox{$\scriptstyle >$}\,}
\def\lsim{\,\lower.25ex\hbox{$\scriptstyle\sim$}\kern-1.30ex%
\raise 0.55ex\hbox{$\scriptstyle <$}\,}
\newcommand{\pomfluxarg}{f_{\pom / p}\,(x_\pom)}
\newcommand{\dsf}{\mbox{$F_2^{D(3)}$}}
\newcommand{\dsfva}{\mbox{$F_2^{D(3)}(\beta,Q^2,x_{I\!\!P})$}}
\newcommand{\dsfvb}{\mbox{$F_2^{D(3)}(\beta,Q^2,x)$}}
\newcommand{\dsfpom}{$F_2^{I\!\!P}$}
\newcommand{\gap}{\stackrel{>}{\sim}}
\newcommand{\lap}{\stackrel{<}{\sim}}
\newcommand{\fem}{$F_2^{em}$}
\newcommand{\tsnmp}{$\tilde{\sigma}_{NC}(e^{\mp})$}
\newcommand{\tsnm}{$\tilde{\sigma}_{NC}(e^-)$}
\newcommand{\tsnp}{$\tilde{\sigma}_{NC}(e^+)$}
\newcommand{\st}{$\star$}
\newcommand{\sst}{$\star \star$}
\newcommand{\ssst}{$\star \star \star$}
\newcommand{\sssst}{$\star \star \star \star$}
\newcommand{\tw}{\theta_W}
\newcommand{\sw}{\sin{\theta_W}}
\newcommand{\cw}{\cos{\theta_W}}
\newcommand{\sww}{\sin^2{\theta_W}}
\newcommand{\cww}{\cos^2{\theta_W}}
\newcommand{\trm}{m_{\perp}}
\newcommand{\trp}{p_{\perp}}
\newcommand{\trmm}{m_{\perp}^2}
\newcommand{\trpp}{p_{\perp}^2}
\newcommand{\alp}{\alpha_s}

\newcommand{\alps}{\alpha_s}
\newcommand{\sqrts}{$\sqrt{s}$}
\newcommand{\LO}{$O(\alpha_s^0)$}
\newcommand{\Oa}{$O(\alpha_s)$}
\newcommand{\Oaa}{$O(\alpha_s^2)$}
\newcommand{\PT}{p_{\perp}}
\newcommand{\JPSI}{J/\psi}
\newcommand{\sh}{\hat{s}}
\newcommand{\uh}{\hat{u}}
\newcommand{\MP}{m_{J/\psi}}
\newcommand{\PO}{I\!\!P}
\newcommand{\xbj}{x}
\newcommand{\xpom}{x_{\PO}}
\newcommand{\ttbs}{\char'134}
\newcommand{\xpomlo}{3\times10^{-4}}  
\newcommand{\xpomup}{0.05}  
\newcommand{\dgr}{^\circ}
\newcommand{\pbarnt}{\,\mbox{{\rm pb$^{-1}$}}}
\newcommand{\gev}{\,\mbox{GeV}}
\newcommand{\WBoson}{\mbox{$W$}}
\newcommand{\fbarn}{\,\mbox{{\rm fb}}}
\newcommand{\fbarnt}{\,\mbox{{\rm fb$^{-1}$}}}
%
%
\newcommand{\qsq}{\ensuremath{Q^2} }
\newcommand{\gevsq}{\ensuremath{\mathrm{GeV}^2} }
\newcommand{\et}{\ensuremath{E_t^*} }
\newcommand{\rap}{\ensuremath{\eta^*} }
\newcommand{\gp}{\ensuremath{\gamma^*}p }
\newcommand{\dsiget}{\ensuremath{{\rm d}\sigma_{ep}/{\rm d}E_t^*} }
\newcommand{\dsigrap}{\ensuremath{{\rm d}\sigma_{ep}/{\rm d}\eta^*} }
\newcommand{\dedx}{\ensuremath{{\rm d} E/{\rm d} x}}
\def\Journal#1#2#3#4{{#1} {\bf #2} (#3) #4}
\def\NCA{Nuovo Cimento}
\def\RPP{Rep. Prog. Phys.}
\def\ARNPS{Ann. Rev. Nucl. Part. Sci.}
\def\NIM{Nucl. Instrum. Methods}
\def\NIMA{{Nucl. Instrum. Methods} {\bf A}}
\def\NPB{{Nucl. Phys.}   {\bf B}}
\def\NPPS{Nucl. Phys. Proc. Suppl.} 
\def\NPPSC{{Nucl. Phys. Proc. Suppl.} {\bf C}}
\def\PR{Phys. Rev.}
\def\PLB{{Phys. Lett.}   {\bf B}}
\def\PRL{Phys. Rev. Lett.}
\def\PRD{{Phys. Rev.}    {\bf D}}
\def\PRC{{Phys. Rev.}    {\bf C}}
\def\ZPC{{Z. Phys.}      {\bf C}}
\def\EJC{{Eur. Phys. J.} {\bf C}}
\def\EPL{{Eur. Phys. Lett.} {\bf}}
\def\CPC{Comp. Phys. Commun.}
\def\NP{{Nucl. Phys.}}
\def\JPG{{J. Phys.} {\bf G}} 
\def\EPC{{Eur. Phys. J.} {\bf C}}
\def\PRSL{{Proc. Roy. Soc.}} {\bf}
\def\PETF{{Pi'sma. Eksp. Teor. Fiz.}} {\bf}
\def\JETPL{{JETP Lett}}{\bf}
\def\IJTP{Int. J. Theor. Phys.}
\def\HJ{Hadronic J.}

\begin{titlepage}
\begin{figure}[!t]
DESY 04--032 \hfill ISSN 0418--9833 \\
March 2004
\end{figure}
\bigskip
\vspace*{2cm}
\begin{center}
\begin{Large}
{\boldmath \bf Measurement of Anti-Deuteron Photoproduction  
 and a Search for Heavy Stable Charged Particles at HERA\\}
\vspace*{2.cm} 
H1 Collaboration \\  
\end{Large}
\end{center}

\vspace*{2cm}

\begin{abstract}
\noindent
The cross section for anti-deuteron photoproduction 
is measured at HERA at a mean centre-of-mass energy of 
$W_{\gamma p} = 200 \gev$ in the range $0.2 < p_T/M < 0.7$ 
and $|y| < 0.4$, where $M$, $p_T$ and $y$ are the mass, transverse
momentum and rapidity in the laboratory frame of the anti-deuteron, 
respectively. The numbers of anti-deuterons per event are found to 
be similar in photoproduction to those in central proton-proton collisions 
at the CERN ISR but much lower than those in 
central Au-Au collisions at RHIC. The coalescence parameter 
$B_2$, which characterizes the likelihood of anti-deuteron production, 
is measured in photoproduction 
to be $0.010 \pm 0.002 \pm 0.001$, which is much higher 
than in Au-Au collisions at a similar nucleon-nucleon centre-of-mass energy. 
No significant production of particles heavier than deuterons is observed 
and upper limits are set on the photoproduction cross sections for such 
particles.  

\end{abstract}

\vspace*{1.5cm}

\begin{center}
(Submitted to European Physical Journal C)
\end{center}

\end{titlepage}

\begin{flushleft}

A.~Aktas$^{10}$,               
V.~Andreev$^{26}$,             
T.~Anthonis$^{4}$,             
A.~Asmone$^{33}$,              
A.~Babaev$^{25}$,              
S.~Backovic$^{37}$,            
J.~B\"ahr$^{37}$,              
P.~Baranov$^{26}$,             
E.~Barrelet$^{30}$,            
W.~Bartel$^{10}$
% ...

\bigskip{\it
% ...
$ ^{4}$ Inter-University Institute for High Energies ULB-VUB, Brussels;
          Universiteit Antwerpen (UIA), Antwerpen; Belgium$^{ c}$ \\
% ...
$ ^{10}$ DESY, Hamburg, Germany \\
% ...
$ ^{25}$ Institute for Theoretical and Experimental Physics,
          Moscow, Russia$^{ l}$ \\
 $ ^{26}$ Lebedev Physical Institute, Moscow, Russia$^{ e}$ \\
% ...
$ ^{30}$ LPNHE, Universit\'{e}s Paris VI and VII, IN2P3-CNRS,
          Paris, France \\
% ...
$ ^{33}$ Dipartimento di Fisica Universit\`a di Roma Tre
          and INFN Roma~3, Roma, Italy \\
% ...
$ ^{37}$ DESY, Zeuthen, Germany \\
% ...
$ ^c$ Supported by FNRS-FWO-Vlaanderen, IISN-IIKW and IWT \\
% ...
$ ^e$ Supported by the Deutsche Forschungsgemeinschaft \\
% ...
$ ^l$ Partially Supported by Russian Foundation
      for Basic Research, grant    no. 00-15-96584 \\
}
\end{flushleft}

\newpage
\section{Introduction}

 This paper describes a measurement of the rate of production 
of anti-deuterons in photon-proton collisions at HERA performed 
by the H1 Collaboration. The measurement is of particular interest in 
the context of recent studies of heavy ion collisions \cite{TS2}. In 
these collisions, the deuteron and anti-deuteron production rate is 
thought to depend on the dimensions of the collision ``fireball'' at the stage 
at which the hadrons decouple \cite{TS1}, i.e. when final state interactions 
become unimportant. This is the so-called ``thermal freeze-out'' region.

Further, a search is performed for the photoproduction of unknown charged   
stable heavy particles in the highest energy electron-proton 
collisions currently accessible in the laboratory. The discovery of such 
particles would be an indication for physics beyond the Standard Model.

The production of nuclei in particle collisions can be described
in terms of the coalescence model. In this model \cite{fitch9},  
the cross section, $\sigma_A$, for the formation of an object with 
$A$ nucleons with total energy $E_A$ and momentum $P$,  
is related to that for the production of free nucleons in the same reaction, 
$\sigma_N$, with energy $E_N$ and momentum $p=P/A$, by  
\begin{equation}
\frac{1}{\sigma} \frac{E_A \rm d^3 \sigma_A}{\rm d^3 P} = B_A \left( \frac{1}{\sigma}\frac{E_N \rm d^3 \sigma_N}{\rm d^3 p} \right) ^A, 
\label{coal}
\end{equation}   
where $B_A$ is the coalescence parameter, which is inversely proportional 
to the source volume in heavy ion collisions \cite{TS1}, and $\sigma$ is 
the total interaction cross section of the colliding particles.   

For the measurements described here, the particles are identified through 
a combination of their specific ionisation energy loss, $\dedx$, and 
their momenta. The numbers of anti-deuterons are measured in the laboratory 
frame in the rapidity region $|y| < 0.4$ which corresponds to a 
centre-of-mass rapidity of between 1.6 and 2.4 units\footnote
{The rapidity and pseudorapidity are defined by 
$y=0.5~\ln [(E+p_z)/(E-p_z)]$ and $\eta = -\ln(\tan\theta/2)$ for a 
particle with total energy $E$, $z$ component of momentum $p_z$ and 
polar angle $\theta$. The $+z$-axis (forward direction) is taken to be 
along the proton beam direction.}. 
In this range, the multiplicity distributions are on the central 
plateau \cite{PDG1} and so comparisons can reasonably be  made with      
the measured numbers of anti-deuterons per event  
in central proton-proton collisions \cite{ISR1,ISR2}. 
Both are contrasted with data from heavy ion collisions \cite{STAR, 
Hansen}. These comparisons are restricted to central collisions  
at a centre-of-mass energy greater than $50 \gev$, i.e. well above the 
threshold for anti-deuteron production. There have been several other 
measurements of anti-deuteron production in proton-proton ($pp$) \cite{pptod}, 
proton-nucleus ($pA$) \cite{pAtod}, nucleus-nucleus ($AA$) \cite{Hansen} 
and electron-positron \cite{ee} collisions. These measurements 
are either for non-central production or are at a centre-of-mass energy 
below $50 \gev$. 
 
\section{Experimental Procedure}
\subsection{The H1 Detector}
Collisions of $27.6 \gev$ positrons with $820 \gev$ protons at HERA 
are detected 
in the H1 detector, which is described in detail elsewhere \cite{fitch5}. 
The components of the detector important in this analysis are 
the small angle positron tagger, the central tracker, the 
backward
Spaghetti-type calorimeter (SpaCal) and the liquid argon (LAr) calorimeter.  

The positron tagger, located at $33\,\hbox{m}$ from the interaction
point in the outgoing positron beam direction, 
is used to trigger on photoproduction events and to measure the energy 
of the scattered positron, from which the total photon-proton 
centre-of-mass energy, 
$W_{\gamma p}$, is deduced.  The central track detector, 
surrounding the 9 cm diameter aluminium beam pipe of thickness 1.7 mm,  
consists of concentric central jet drift chambers (CJCs) with inner (CJC1) 
and outer (CJC2) chambers and two additional drift chambers which measure 
the $z$ coordinates of tracks. The pseudorapidity range covered by the 
central track detector is $|\eta|<1.5$. The CJC has 56 sensitive wire 
layers: 24 in CJC1 covering radii from 20.3 cm to 45.1 cm and 32 in CJC2 
covering radii from 53.0 cm to 84.4 cm. 
The detector is placed inside a 
uniform magnetic field of 1.15~T, allowing measurements of 
the track transverse momentum to be made with a resolution of 
$\sigma_{p_T}/p_T \approx 0.009\cdot p_T [\gev] \oplus 0.015 $.  
The specific energy loss, $\dedx$, of the charged particles is 
also measured in this detector with a resolution 
$\sigma (\dedx)/(\dedx)$ of $7.5 \% $ for 56
hits on a minimum ionising track. The LAr calorimeter covers the 
angular range $4^\circ < \theta < 154^\circ$ with the forward 
region defined to be  $4^\circ < \theta < 25^\circ$.  
The SpaCal calorimeter covers the backward region, i.e. 
the angular range $153^\circ < \theta < 177.8^\circ$. 
The luminosity is measured via the well understood Bethe-Heitler 
process, $ep \rightarrow ep\gamma$, using a photon detector at $0^\circ$ 
to the positron beam direction.   

\subsection{Trigger Conditions, Event and Track Selection}
The measurements presented here are based on H1 data taken with
minimally biased triggers in 1996 and correspond to an 
integrated luminosity of $5.53 \pm 0.11$ pb$^{-1}$. 
Photoproduction events are triggered by requiring the presence of 
tracks in the CJC and of a scattered positron in the positron tagger, 
which ensures that the photon virtuality $Q^2<10^{-2}~\gevsq$.  
The following selection criteria are applied in order to reduce the 
background contamination and to ensure good reconstruction of the event 
kinematics.  The selected events are required to lie within the interval
$165 < W_{\gamma p} < 252 \gev$ (average $\langle W_{\gamma p} \rangle 
=200 \gev$).  In this range the total acceptance of the positron tagger 
for photoproduction events is $0.46 \pm 0.02$. In addition, five or 
more tracks are required to be reconstructed in the CJC. These tracks 
must point to a common vertex with $z$ coordinate within $\pm 30$~cm 
of the nominal interaction point.  


Candidate tracks for particle identification are selected in the range 
$|\eta | < 1$ so that they 
are well contained within the CJC. Here the track reconstruction and 
particle identification efficiencies are high. 
Two track selection schemes are used: the ``hard'' and ``soft'' selections.
The minimum ionising particle (MIP) background is larger in the soft 
selection scheme than in the hard selection scheme. The 
hard selection is used to search for rare heavy particles, when it is 
necessary to 
minimise this background and to have optimum $\dedx$ resolution. 
The soft selection is used only for copiously produced particles such as 
protons and anti-protons.  

In the soft selection, the tracks are required to have at least 10 hits 
and to have a start point at radius $< 30$ cm from the beam line and 
an end point at  radius $> 37.5$ cm. The total measured radial track length 
is required to be more than 10 cm and the specific energy loss to be 
more than twice that of a MIP,   
i.e. $\log_{10} \dedx > 0.3$.\footnote{Throughout the paper $\dedx$ is 
given as the ratio of the specific energy loss of the track to that of a MIP.}
These criteria are loose enough to ensure that the combined 
track reconstruction and particle 
identification efficiency is high (measured to be $98.8 \pm 0.2\%$).  
In the hard selection, tracks are required to pass through both CJC1 
and CJC2.  The number of hits in CJC1 is required to be at least 20  
with a minimum of 40 hits in both chambers. The total measured radial 
track length must be greater than 35 cm. In addition, the selected 
tracks must have a minimum of 75$\%$ of the total number of possible 
hits and a specific energy loss of more than 2.5 times that of a MIP,   
i.e. $\log_{10} (\dedx) > 0.4$. 

\subsection{Particle Identification.}
\label{IDP}

The mass, $M$, associated with each charged particle track is deduced 
from the track momentum, $p$, and the 
most probable specific energy loss, $\dedx_0$, which is determined 
using a Bayesian log-likelihood method \cite{fitch7,WalterGigi}. The 
value of $\dedx_0$ is chosen for each track such that the
likelihood function  
\begin{equation}
 \log L = \sum_i^N \log P(\dedx_i|\dedx_0)
\label{LH}
\end{equation}
is maximised. 
Here, $P(\dedx_i|\dedx_0)$ is the probability 
that the $i^{\rm th}$ measured value of $\dedx$ results from a particle with 
most probable specific energy loss $\dedx_0$ which is treated as a variable 
in equation \ref{LH}.  This probability is computed from a 
parameterisation of the Landau distribution. The ratio $p/M$ is then 
obtained in an iterative way from $\dedx_0$, assuming that the particle 
is singly charged, using a parameterisation of the Bethe-Bloch formula 
for the restricted energy loss \cite{PDG} which includes corrections 
for apparatus effects. 
The value of $M$ is then calculated using the measured track momentum.  
The maximum measurable mass using this technique is beyond the  
limit set by the centre of mass energy.  

\begin{figure}[htb]
\begin{center}
\hspace*{-3mm}\epsfig{file=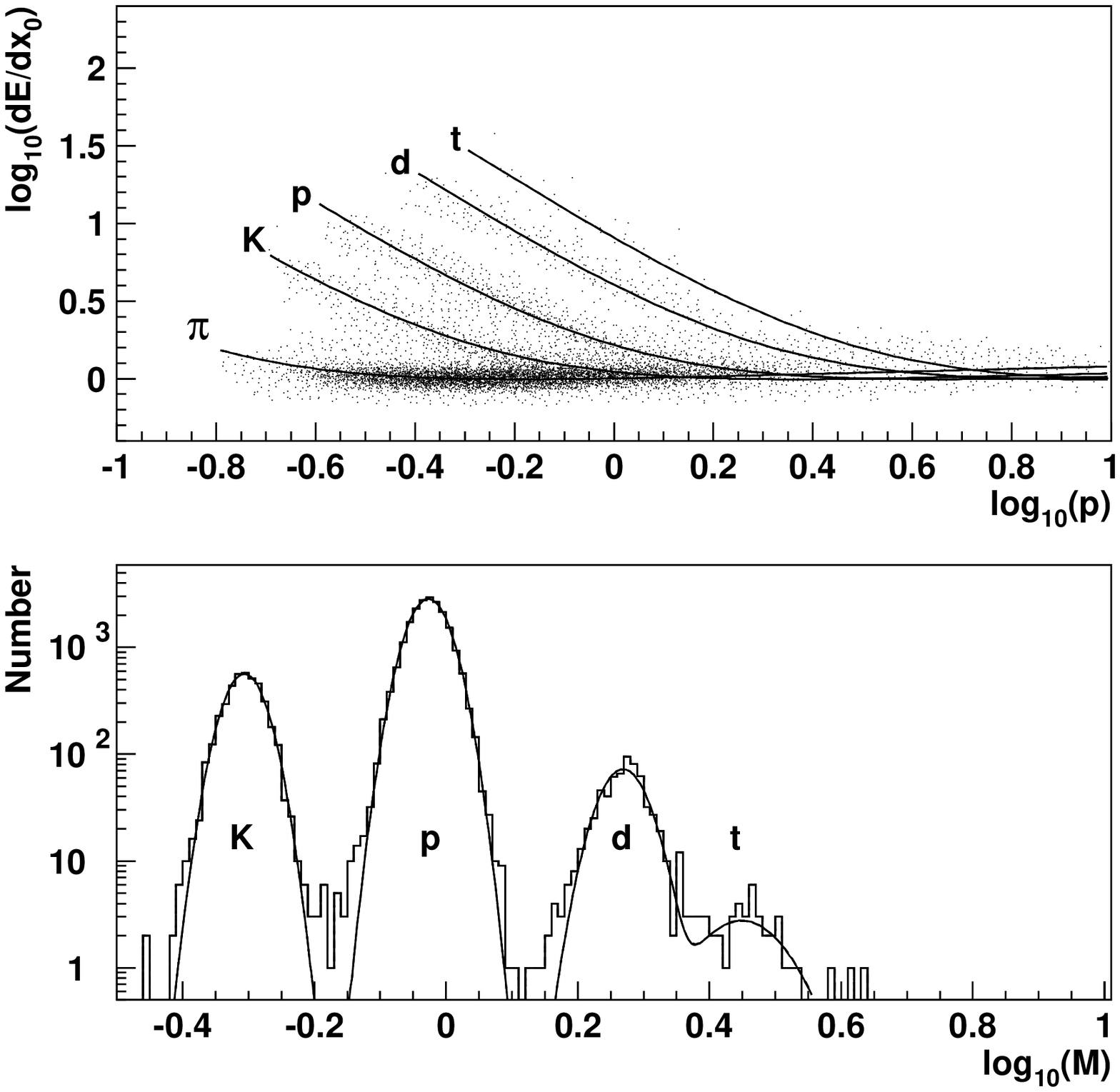,
bburx=582,bbury=715,bbllx=12,bblly=124,
height=19.0cm,angle=0}
\end{center} 
\vspace*{-16mm}
\caption{ Upper plot - the observed specific ionisation energy 
loss, $\dedx$, (normalised to that from a MIP), obtained from the 
log-likelihood method, versus track momentum (in GeV) for a sample 
of positively charged 
tracks from the hard selection (before application of the specific 
energy loss cut, see text). The smooth curves 
show the expected mean specific energy loss for the 
different particle species. Lower plot - the spectrum of masses  
(in $\gev$) for $\log_{10} \dedx > 0.4$,  
deduced as described in the text. The smooth curves 
represent Gaussian fits to each peak.}
\label{exfig1} 
\end{figure}

Figure \ref{exfig1} (upper plot) shows the specific energy loss for 
positively charged tracks, determined in this way, plotted against the 
track momenta. Clear bands can be seen corresponding to 
pions, kaons, protons, deuterons and tritons. 
The lower plot shows the spectrum 
of masses assigned to the tracks using the procedure described 
above. The smooth curves show parabolic fits to the $\log_{10} M$ 
distributions (i.e. Gaussians) in each of the different mass peaks. 
Some deviations from 
Gaussian behaviour in the tails of the distributions are observed. The 
resolution is $\delta M/M \sim 7\%$ as determined from the widths of the 
Gaussian fits. The log-likelihood method 
adopted here is found to have 
better mass resolution and leads to a 
more Gaussian-like distribution than the method used previously by 
H1 \cite{Steinhart}. The particle type for each track is identified as 
the mass of the closest known particle.

\subsection{Background Determination}
\label{bckgd}

The distributions of the $z$ vertex coordinate and the distance of 
closest approach (DCA) of the tracks to the beam line in the transverse 
plane are used to distinguish the tracks produced in photoproduction from 
those produced by interactions of the beams with residual gas in the 
beam pipe, termed beam-gas interactions, or 
from the secondary interactions of photoproduced particles in the material 
of the beam pipe or the detector, termed material background. 
The beam-gas backgrounds are labelled $pG$ for proton-gas and $eG$ for 
electron-gas interactions.  

\begin{figure}[htb]
\begin{center}
\hspace*{-2mm}\epsfig{file=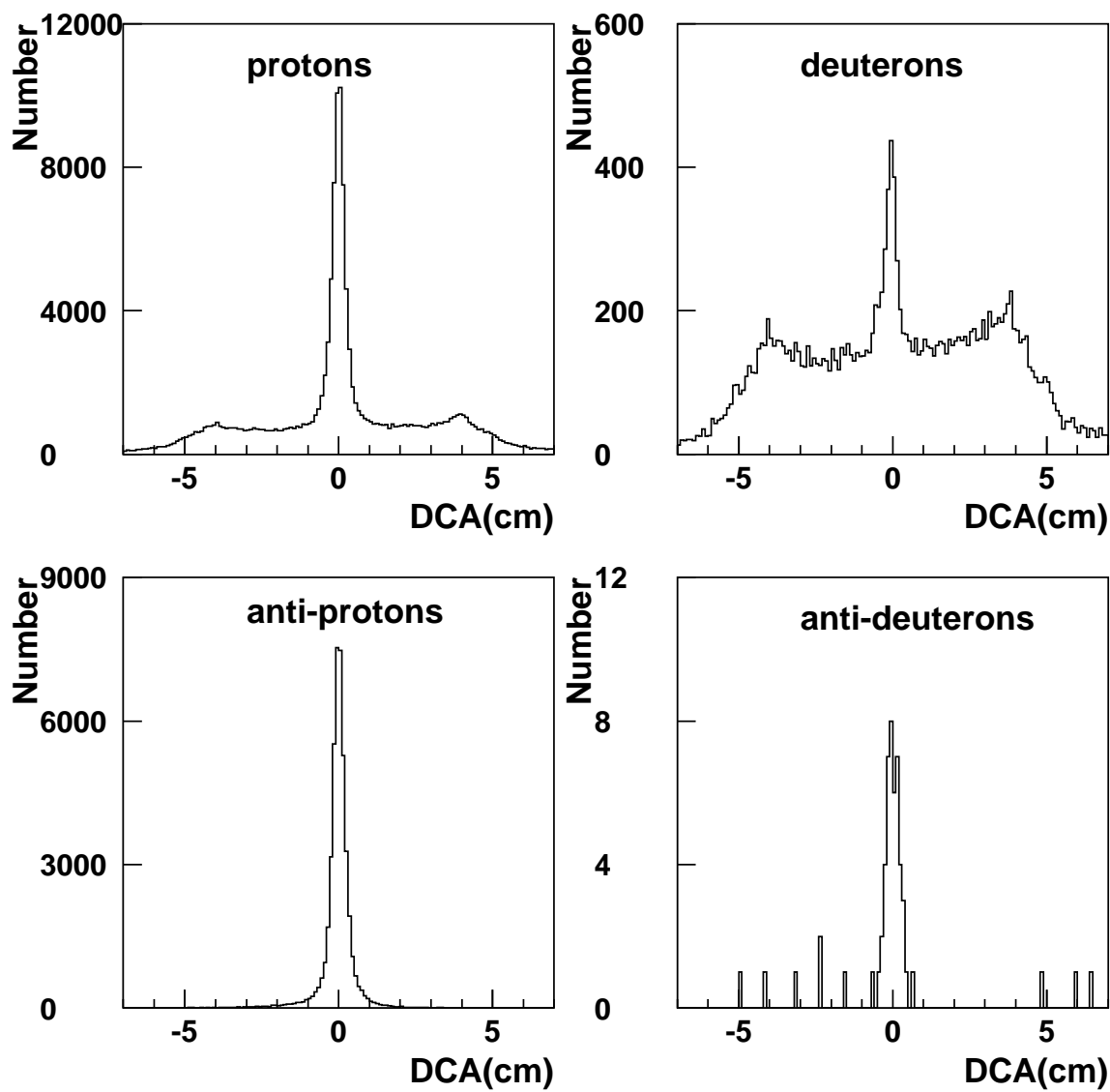,
bburx=582,bbury=715,bbllx=12,bblly=124,
height=18.0cm,angle=0}
\end{center}
\vspace*{-14mm} 
\caption{ The distribution of the distance of closest approach (DCA)
of the track to the event vertex in the plane transverse to the beam
direction for $p$, $d$, $\bar p$ and $\bar d$ candidates    
with momenta larger than $0.5 \gev$ from the hard selection and before
any track timing cuts. The peak at zero from beam induced events sits 
on the material background which is much smaller for $\bar p$, 
$\bar d$ than for $p$, $d$. }
\label{exfig2} 
\end{figure}

Figure \ref{exfig2} shows the DCA distributions. Tracks from interactions 
of the beam particles contribute to the peaks at zero, whereas the material 
background gives rise to the observed smooth background distributions. 
The material background for positive tracks results mainly from 
secondary interactions with the material of the beam pipe or the other  
material before the  CJC sensitive region.  
This can be seen in the upper plots of figure~\ref{exfig2} where it 
leads to an approximately flat background with small peaks at DCA $\sim 4$ cm.
For negative tracks, the material background is much 
smaller (see figure \ref{exfig2} lower plots). It arises  
mainly from protons and deuterons which are back-scattered from the 
calorimeters into the CJC. These albedo particles, which lie outside the main 
peak in figure \ref{exfig2}, are delayed due to 
their extra distance of travel by times of about $6$ ns 
relative to the arrival time of tracks coming directly from the 
photoproduction interaction vertex, as measured in the CJC. 
The selection of negative particles with masses greater than the proton 
mass is supplemented by the requirement that this delay be less than 
4 ns, which is observed to reduce such backgrounds by about a factor of two. 

The number of particles corrected for the material background is obtained 
by subtracting the number in the sidebands of the DCA distribution 
($1.5 < |\textrm{DCA}| < 3.0$ cm), normalised to the width of the selected 
region around the peak, from the total number of particles in the peak region. 
This width is chosen to be $\pm 1.5$ cm for $p$ and $\bar p$ and 
$\pm 0.5$ cm for heavier particles for which the DCA resolution is better.    

The beam gas background is measured most accurately by dividing the data into 
four event samples depending on the presence or absence of energy 
in the forward part of the LAr or in the backward direction in the SpaCal. 
The event sample with both forward and backward 
energy (labelled C$_{11}$ and comprising $89.4\%$ of the total), arises     
dominantly from photoproduction ($\gamma p$) which produces a roughly 
uniform distribution of energy in the apparatus for the $W_{\gamma p}$ 
range of this measurement. The event sample 
with forward but no backward energy (labelled C$_{10}$ and comprising 
$8.5\%$ of the total) arises dominantly from $pG$ interactions, since protons 
interacting in the interaction region with a nearly stationary target 
produce mainly forward but little significant backward energy. The event 
sample with backward but no forward energy 
(labelled C$_{01}$ and comprising  $2.0\%$ of the total), is enriched in 
$eG$ interactions for similar reasons. Only $0.1\%$ of the events have 
neither forward nor backward calorimetric energy (sample C$_{00}$).
 
\begin{figure}[htb]
\begin{center}
\hspace*{-3mm}\epsfig{file=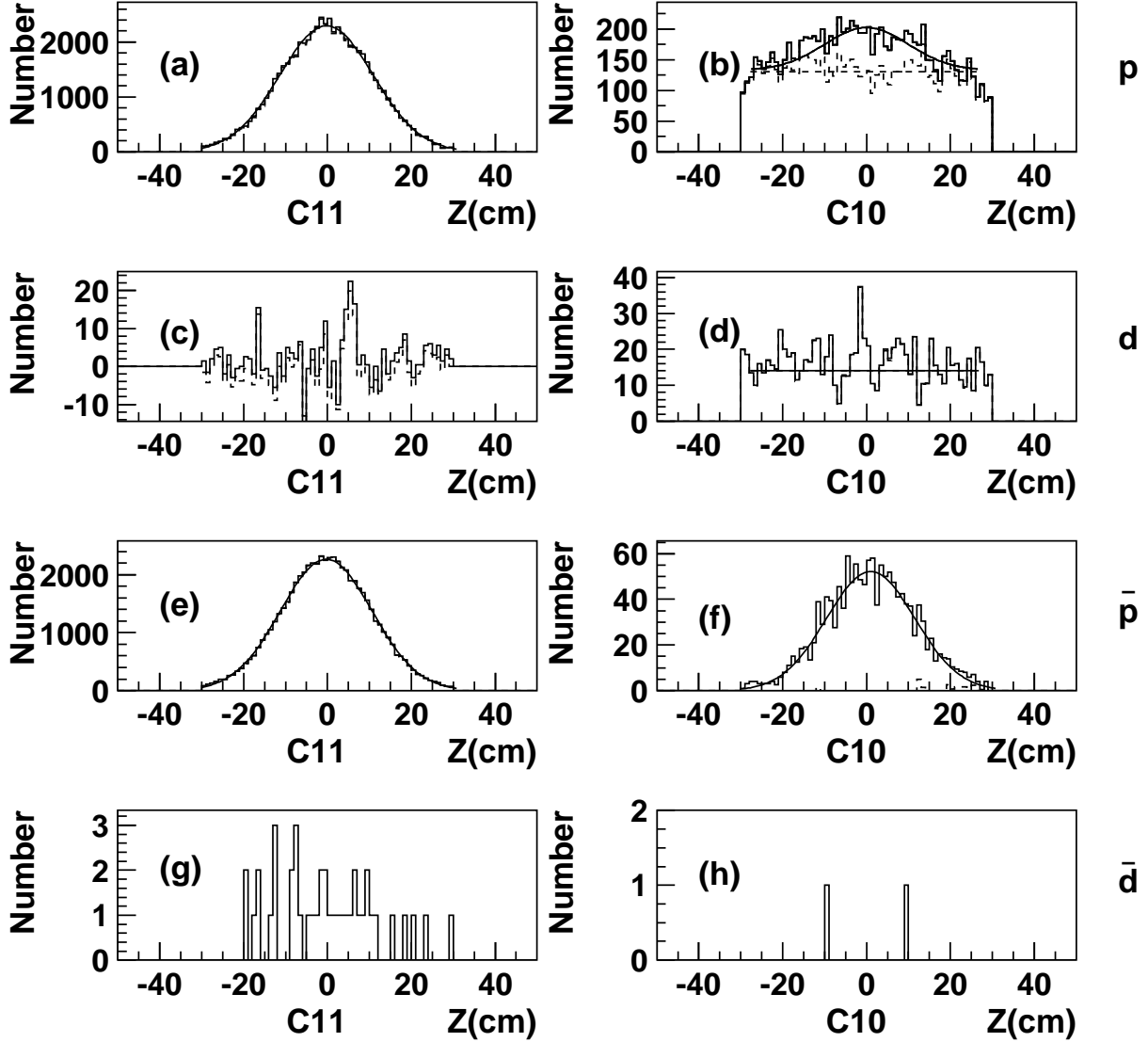,
bburx=582,bbury=715,bbllx=12,bblly=124,
height=18.0cm,angle=0}
\end{center} 
\vspace*{-15mm}
\caption{The $z$ vertex distributions for 
identified protons (a,b), deuterons (c,d), anti-protons (e,f) 
and anti-deuterons (g,h) in the hard 
selected samples after correction for the material backgrounds. The 
C$_{11}$ samples (figures a,c,e,g) are dominated by photoproduction events 
and the C$_{10}$ sample (figures b,d,f) are dominated by $pG$ events. The 
solid curves (figures a,b,c,d) show the fits of a Gaussian distribution 
(expected from photoproduction) and a linear background 
(expected from $pG$ interactions). The lack of background and the 
relatively small number of events in (f) show that there is little 
$\bar p$ production in $pG$ events. The number of $\bar d$ events in 
(h) is compatible with that expected from the $\gamma p$ contamination 
of C$_{10}$ sample. The lack 
of background in (a) and (e) shows that there are few $pG$ events in 
the C$_{11}$ sample. The dashed histograms represent the background after 
subtraction of the fitted (b) or calculated (c,f) Gaussians.}
\label{exfig5} 
\end{figure}

The data are separated into the $\gamma p$, $pG$ and $eG$ components 
in each sample, C$_{ik}$, by studying the $z$ vertex distributions (see 
figure \ref{exfig5}). 
The following procedure is adopted for the separation, the results of 
which are shown in table \ref{Table1} for 
the hard selected tracks. First, the number of $\gamma p$ and $pG$ ($eG$) 
tracks in the samples C$_{10}$ (C$_{01}$), for each particle type, is 
measured by fitting the $z$ vertex distributions to the sum of a Gaussian 
and a linear background. 
The integral of the Gaussian is taken to be the number 
of particles from $\gamma p$ interactions, 
$N_{10}^{\gamma p}$ ($N_{01}^{\gamma p}$), while the linear background 
determines the number of $pG~(eG)$ tracks, $N_{10}^{pG}$ ($N_{01}^{eG}$),  
in the samples. 
Here the number of $eG$ ($pG$) events in the C$_{10}$ (C$_{01}$) sample  
is neglected. The number of $eG$ particles in the C$_{11}$ sample is also 
negligible.
When the distribution has no visible Gaussian shape (e.g. 
figure \ref{exfig5}d), the value
of $N_{ik}^{\gamma p}$ is so low that a measurement is impossible (the 
dashes in table \ref{Table1}). 
Second, the number of $pG$ particles in each C$_{11}$ sample, $N_{11}^{pG}$, 
is obtained by assuming that the probability 
that a true $pG$ event has significant backward energy is small.
Hence, the probability that a $pG$ event appears in the C$_{11}$ sample 
is approximately the probability that random noise above threshold 
occurs in the SpaCal 
calorimeter in a $pG$ event and this should be independent of particle type.  
This probability is measured from the ratio of the numbers of events 
in the C$_{11}$ and C$_{10}$ samples for pure $pG$ events, i.e. samples 
which should have little contribution from photoproduction. Deuterons 
with $z$ vertex $|z|>20$~cm and $\cos \theta > 0$ 
are used for this, as is an independent sample of events with two 
identified protons and no identified anti-proton, which also shows no 
significant Gaussian shape in the $z$ vertex distribution.  The two 
independent measurements give ratios which agree within errors, confirming 
the assumption, with a mean of $0.151\pm 0.045$. 
The quantity $N_{11}^{pG}$ is obtained by multiplying 
$N_{10}^{pG}$ by this ratio. Finally, the number of $\gamma p$ particles in 
the C$_{11}$ sample, $N_{11}^{\gamma p}$, is obtained by subtracting 
$N_{11}^{pG}$ from the total for this sample, $N_{11}$. The separation 
into components for the C$_{00}$ sample is made using the probabilities 
($\sim 0.02$, estimated from the numbers in table \ref{Table1}) that the 
photoproduced events have zero energy in the forward and backward 
calorimeters. 

\subsection{Track Efficiencies}
\label{effys}

The apparatus is fully sensitive in the range $0.2 < p_T/M < 0.7$ 
and $|y| < 0.4$.  
In order to derive cross sections, corrections for track 
efficiencies (labelled $\epsilon$) must be applied. These are shown 
in detail for anti-deuterons in 
table \ref{dbex} in the measurement intervals of $p_T/M$.  
The efficiencies for anti-protons tend to be somewhat larger than those 
for anti-deuterons, since the soft selection 
is used and the secondary interaction cross section is smaller.  
The efficiencies listed in table \ref{dbex} are defined as follows.
 
\begin{itemize}
\item 
$\epsilon_{\dedx}$ is the correction for migrations across the limit  
$\log \dedx > 0.4$. This is assessed by studying the migrations of 
anti-protons, selected with the looser criterion $\log \dedx > 0.3$. 

\item
$\epsilon_\phi$ is a correction for a region of 
inefficiency in the CJC which developed during the data taking. 

\item
$\epsilon_{cut}$ is a correction for the loss of events outside the 
mass and DCA windows. 

\item
$\epsilon_\sigma$ represents the corrections for the 
losses due to interactions in the material of the apparatus. This is 
extracted from the data using soft selected, identified tracks. 
The number of such tracks which are observed to interact in the material 
between CJC1 and CJC2 is extrapolated to account for the material between 
the interaction point and the CJC1. The corrections agree with estimates 
from known cross sections. 

\item
$\epsilon_{hit}$ represents  
the track reconstruction efficiency which is determined  
by measuring the fraction of soft selected deuterons which enter 
the hard selection sample.

\item
$\epsilon_{trig}$ is the trigger efficiency which  
is determined by two independent methods~ for events containing anti-protons.  
These are assumed to have the same trigger efficiency as anti-deuterons. 
One method involves Monte Carlo studies and the other method uses 
comparisons of the number of events from the main trigger with those 
found by an independent monitor trigger. The two methods give consistent 
results. 

\item
$\epsilon_{tag}$ is the positron tagger acceptance. 

\item 
$\epsilon_{Nch}$ is the correction for the loss of events due to the 
requirement that there be five or more tracks in each event. This is 
deduced by applying the known KNO scaling distribution \cite{KNO}, 
using a sample of events containing anti-protons which has an observed 
mean track multiplicity which matches that for events containing 
anti-deuterons.  

\item
$\epsilon_{t0}$ represents the 
correction for the losses of anti-deuterons outside the defined track 
timing interval. 

\item
$\epsilon_{PhSp}$ is an estimate of the fraction of anti-deuterons lost 
outside the measurement region defined by the limits of 
$|\eta | < 1.0$ and $|y| < 0.4$. 

\end{itemize}

The final row of table \ref{dbex} gives the measured value of the 
differential cross section 
\begin{equation}
\frac{\rm d\sigma}{\rm d(p_T/M)} = \frac{N}{\Delta (p_T/M) {\cal L} F}.
\label{diffsig}
\end{equation}
where $N$ is the number of events in each measurement interval, 
$\Delta(p_T/M)$, corrected for all the efficiencies given in table \ref{dbex}, 
${\cal L}=5.53 \pm 0.11$ pb$^{-1}$ is the integrated luminosity and 
$F = 0.0136$ is the virtual photon flux per incident positron (for 
the details of the calculation see \cite{H1sigT}). 
The first error quoted for the differential cross sections is statistical 
while the second is systematic, where 
the latter arises from the uncertainties in the efficiencies.  
All cross sections are quoted at the bin centres.

\section{Results}

\subsection{Search for Heavy Particles}

\begin{figure}[htb]
\begin{center}
\hspace*{-3mm}\epsfig{file=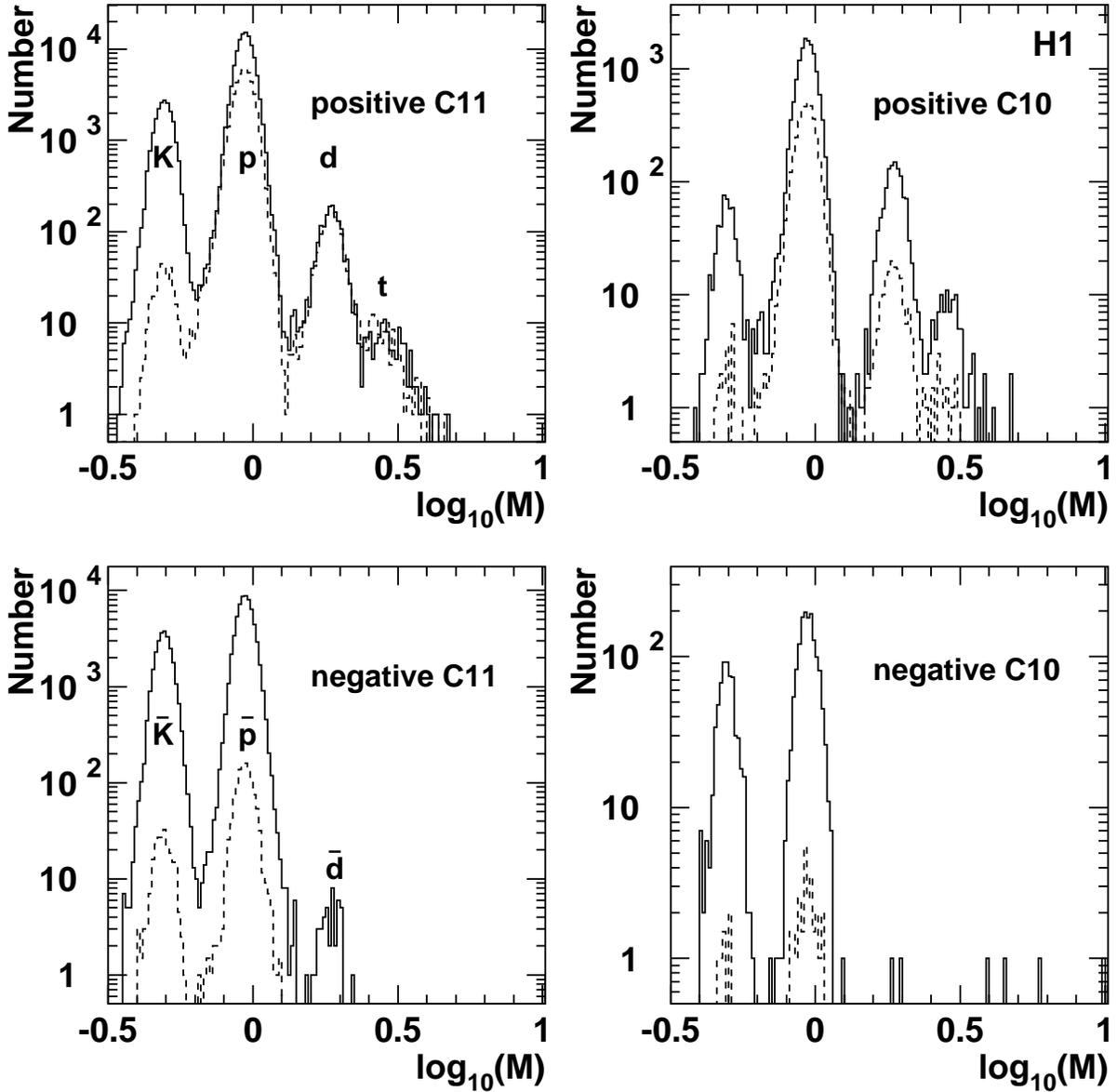,
bburx=582,bbury=715,bbllx=12,bblly=124,
height=19.0cm,angle=0}
\end{center}
\vspace*{-10mm} 
\caption{The mass spectra (with $M$ in $\gev$) for positive and negative  
particles in the hard selected sample for the mainly photoproduction 
(C$_{11}$) and mainly $pG$ (C$_{10}$) samples. The dashed lines show the 
material background deduced from the sideband subtraction method.}
\label{exfig13}
\end{figure}

The observed particle mass spectra are shown in figure \ref{exfig13}. The 
dashed curves show the material backgrounds deduced from the sideband 
subtraction method described in section \ref{IDP}. The mainly $pG$ 
sample, C$_{10}$, contains 6 tracks which have reconstructed masses of 
more than $3~\gev$. They each have specific energy losses 
which are approximately twice that expected for a MIP.  
Visual inspection shows that these are overlapping tracks, probably 
due to relativistic particles which have been merged by 
the pattern recognition software, since the two tracks become visible 
at the ends of their trajectories. No such 
tracks are seen in the dominantly photoproduction sample, C$_{11}$, 
which contains many more events. 
Since they are dominated by background, the C$_{10}$, C$_{01}$ and C$_{00}$ 
samples are omitted from the search for heavy particles in 
photoproduction. 

The observed deuterons and tritons (see table \ref{Table1} and 
figure \ref{exfig13}) are dominantly from the material background and 
upper limits on their photoproduction cross sections are derived from 
the observed numbers of events. These upper limits at the $95\%$ confidence 
level, in the measurement range defined in section \ref{effys}, are 
deduced to be 6.8 and 1.0 nb, respectively.  
No negative particles heavier than anti-deuterons 
and no positive particles heavier than tritons are observed. This 
allows an upper limit of 0.19 nb at the $95\%$ confidence 
level to be set on the photoproduction cross section for any such 
particle type in the same kinematic range. These cross section  
limits are derived assuming the same efficiencies as those for 
anti-deuterons given below.    

\subsection{The Anti-deuteron Cross Section}

\label{sigmas}

A clear signal is seen in figure \ref{exfig13}, consisting of a total 
of 45 anti-deuterons with an estimated material background of $1.0 \pm 0.5$. 
The inclusive cross sections are measured using the 35 particles in 
the sensitive range defined in section \ref{effys} for all samples combined. 
Only two of the 
anti-deuterons are not in the C$_{11}$ sample (both in C$_{10}$).  
This number is compatible with that expected from the probability 
that a photoproduction event has small backward energy ($\sim 2\%$). 

The total cross section for anti-deuteron production is found to be 
$2.7 \pm 0.5 \pm 0.2$ nb in the kinematic range defined in section 
\ref{effys}, by summing all the differential 
cross sections in the final row of table \ref{dbex}. The ratio of the 
number of anti-deuterons to anti-protons in the range $|y| < 0.4$ and 
$0.3 < p_T/M < 0.7$ is measured to be $(5.0 \pm 1.0 \pm 0.5) \cdot 10^{-4}$. 
The lower limit in $p_T/M$ is higher for the ratio than for 
the anti-deuteron cross section measurement    
in order to avoid the uncertainties associated with the large corrections 
to the anti-proton rates at low momentum.    

\begin{figure}
\begin{center}
\epsfig{file=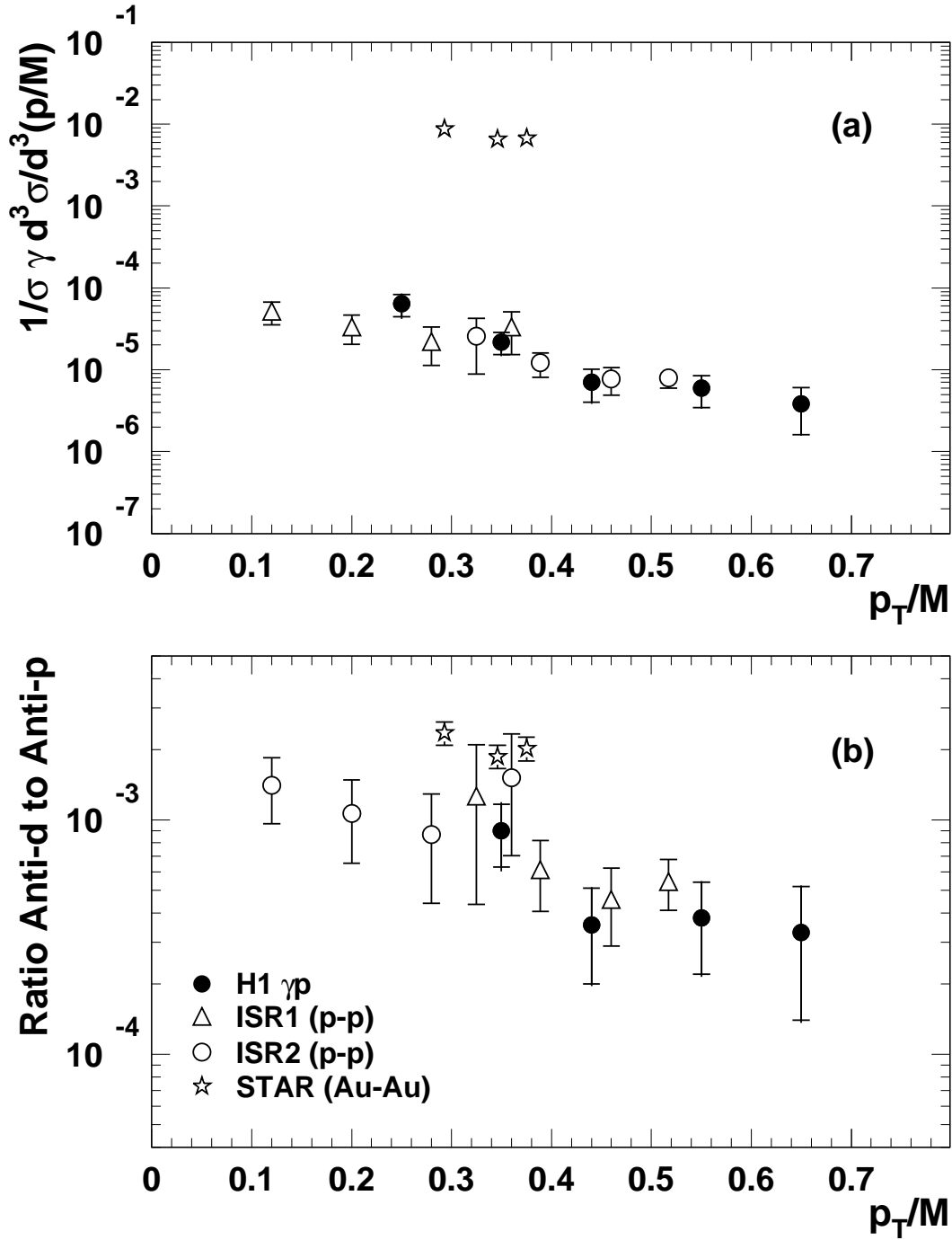,
height=20.0cm,angle=0}
\end{center}
\vspace*{-12mm}
\caption{a) The measured invariant cross section (normalised to the 
relevant total cross section) for inclusive $\bar d$ 
production, compared with the $pp$ and Au-Au data.
b) The measured $\bar d$ to $\bar p$ production ratio as a 
function of $p_T/M$ (solid points) compared with $pp$ data from 
the ISR at $\sqrt s= 53 \gev$ in the central region \cite{ISR1,ISR2} 
and RHIC data on Au-Au collisions \cite{STAR}. The inner error bars 
on the H1 data indicate the statistical and the outer the total 
uncertainties.}
\label{wPT}
\end{figure}

The inclusive anti-deuteron invariant cross section is given by: 
\begin{equation}
\label{siginv}
\gamma \frac{\rm d^3 \sigma}{\rm d^3(p/M)} = M^2 E \frac{\rm d^3 \sigma}{\rm d^3p}=\frac{1}{2 \pi~(p_T/M)~\Delta y}\frac{\rm d\sigma}{\rm d(p_T/M)}
\end{equation}
where $\gamma = E/M$, $\rm d\sigma/\rm d(p_T/M)$ are the differential 
cross sections given in the final row of table \ref{dbex} and  
$\Delta y = 0.8$ is the rapidity range of the measurement.  
Figure \ref{wPT} and table \ref{Results} show the measurements of this 
invariant cross section, normalised to the relevant total cross 
section, taken to be $165 \pm 11 \mu$b for photoproduction \cite{H1sigT}, 
as a function of $p_T/M$.  Figure \ref{wPT} also shows the measured ratio 
of the corrected numbers of anti-deuterons to anti-protons versus $p_T/M$.    

\subsection{Comparison with Other Measurements}

Measurements of the normalised invariant anti-deuteron cross section and 
the ratio of the anti-deuteron to anti-proton production rates performed 
by other high energy experiments in the central region are also shown in 
figure \ref{wPT}. The normalised cross section and ratio measurements 
obtained in $pp$ collisions \cite{ISR1,ISR2} and the photoproduction 
results described here are in good ~agreement, suggesting that the 
processes whereby anti-deuterons are formed are similar in $pp$ and 
$\gamma p$ interactions. The data on Au-Au collisions show a ratio 
which is slightly larger than that in $pp$ collisions and photoproduction. 
However, the cross section for anti-deuteron production normalised 
to the total cross section in Au-Au collisions is over two 
orders of magnitude larger than that in photoproduction or
$pp$ collisions, reflecting the more copious production of
anti-nucleons in heavy ion collisions.

The coalescence parameter $B_2$ is derived by rearranging equation \ref{coal}
to obtain
\begin{equation}
B_2 = \frac {{\frac{1}{\sigma}}{\frac{E_{\bar d} \rm d^3 \sigma_{\bar d}}{\rm d^3 P}}}{
\left(\frac{1}{\sigma}\frac{E_{\bar p} \rm d^3 \sigma_{\bar p}}{\rm d^3 p} \right)^2} 
= \frac{M_{\bar p}^4}{M_{\bar d}^2} \frac{R^2}{\left(\frac{1}{\sigma}\frac{\gamma_{\bar d} \rm d^3 \sigma_{\bar d}}{\rm d^3 (P/M_{\bar d})} \right)}. 
\label{coal1}
\end{equation}
Here, $M_{\bar p}$ and $M_{\bar d}$ are the masses of the $\bar p$ 
and $\bar d$ and $\sigma$, $\sigma_{\bar p}$ and $\sigma_{\bar d}$ 
are the total interaction cross section and 
the partial cross sections for $\bar p$ and $\bar d$ production, respectively, 
as defined in equation \ref{coal} 
with $A=2$ for $\bar d$ production. The quantity, $R$, is 
the measured ratio of the number of anti-deuterons to anti-protons from direct 
production in each bin of $p_T/M$, corrected for anti-protons  
formed remotely from the source by weak decays.      
In the coalescence model, only anti-nucleons produced directly from the 
source can form anti-deuterons. The number of $\bar p$ from direct 
production is taken to be $78 \pm 8 \%$ of the 
number observed. This is estimated using the PYTHIA 
Monte Carlo \cite{PYTH}, the accuracy being determined by the uncertainty in 
the strangeness suppression factor in this model.    
Hence, there is an overall theoretical uncertainty of about 
$20\%$ in the 
determination of $B_2$. The weak decay correction is  
somewhat larger for heavy ion collisions \cite{STAR} which is 
thought to be due to enhanced strangeness production \cite{DH}.

\begin{figure}
\begin{center}
\epsfig{file=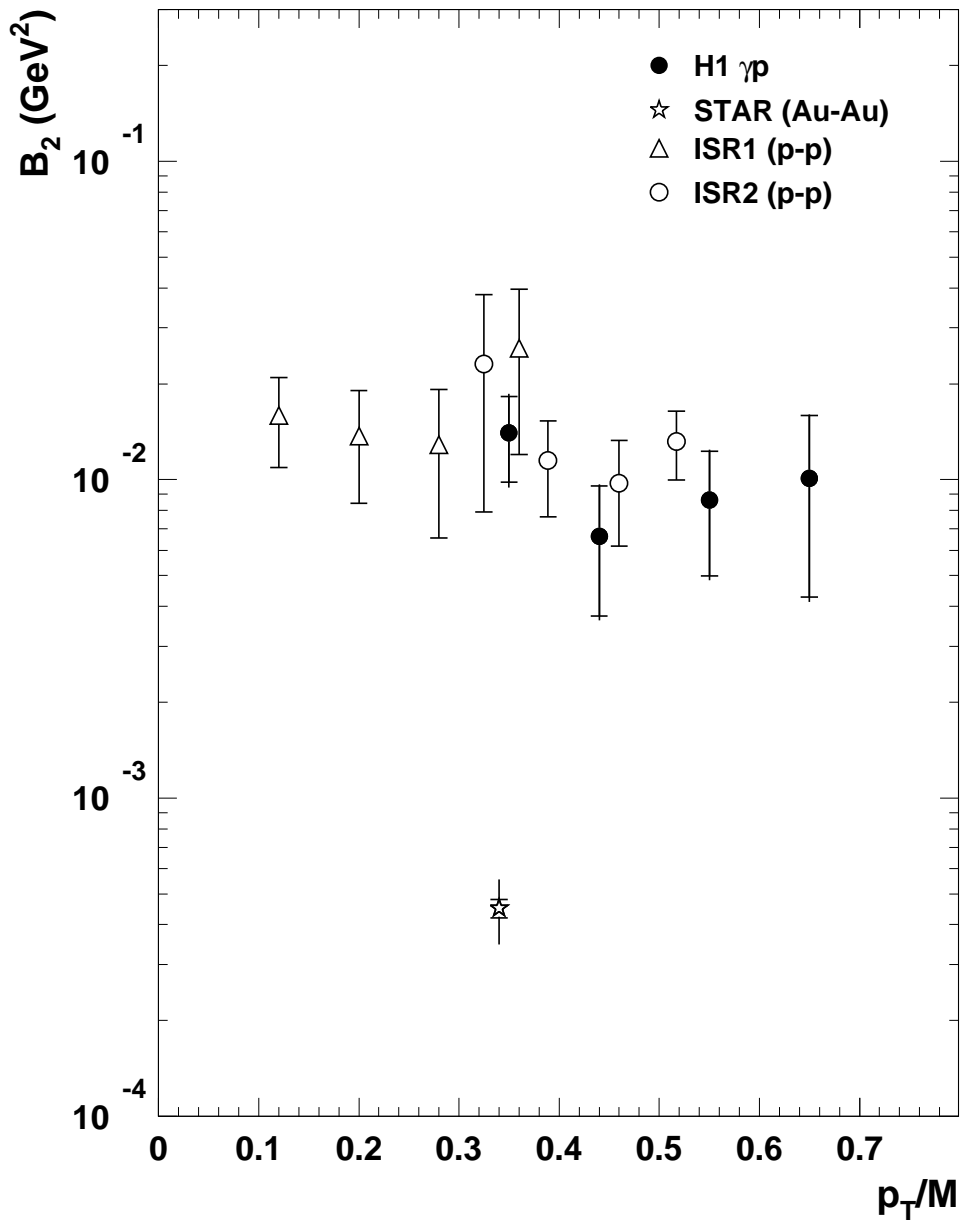,
height=18.0cm,angle=0}
\vspace*{-7mm} 
\end{center}
\caption{The $p_T/M$ dependence of the parameter $B_2$ for $\gamma p$ 
(this experiment), $pp$ \cite{ISR1,ISR2} and Au-Au interactions 
\cite{STAR}. The values of $B_2$ for the $pp$ data are  
deduced from the measured cross sections \cite{ISR1,ISR2}. 
There is a theoretical uncertainty of about $20\%$ (not shown)  
in the determination of $B_2$ for the photoproduction and $pp$ data,  
which arises from the calculation of the fraction of anti-protons produced 
directly (see text). The inner error bars 
on the H1 data indicate the statistical and the outer the total 
uncertainties.}
\label{b2}
\end{figure}

\begin{figure}[htb] 
\begin{center}
\epsfig{file=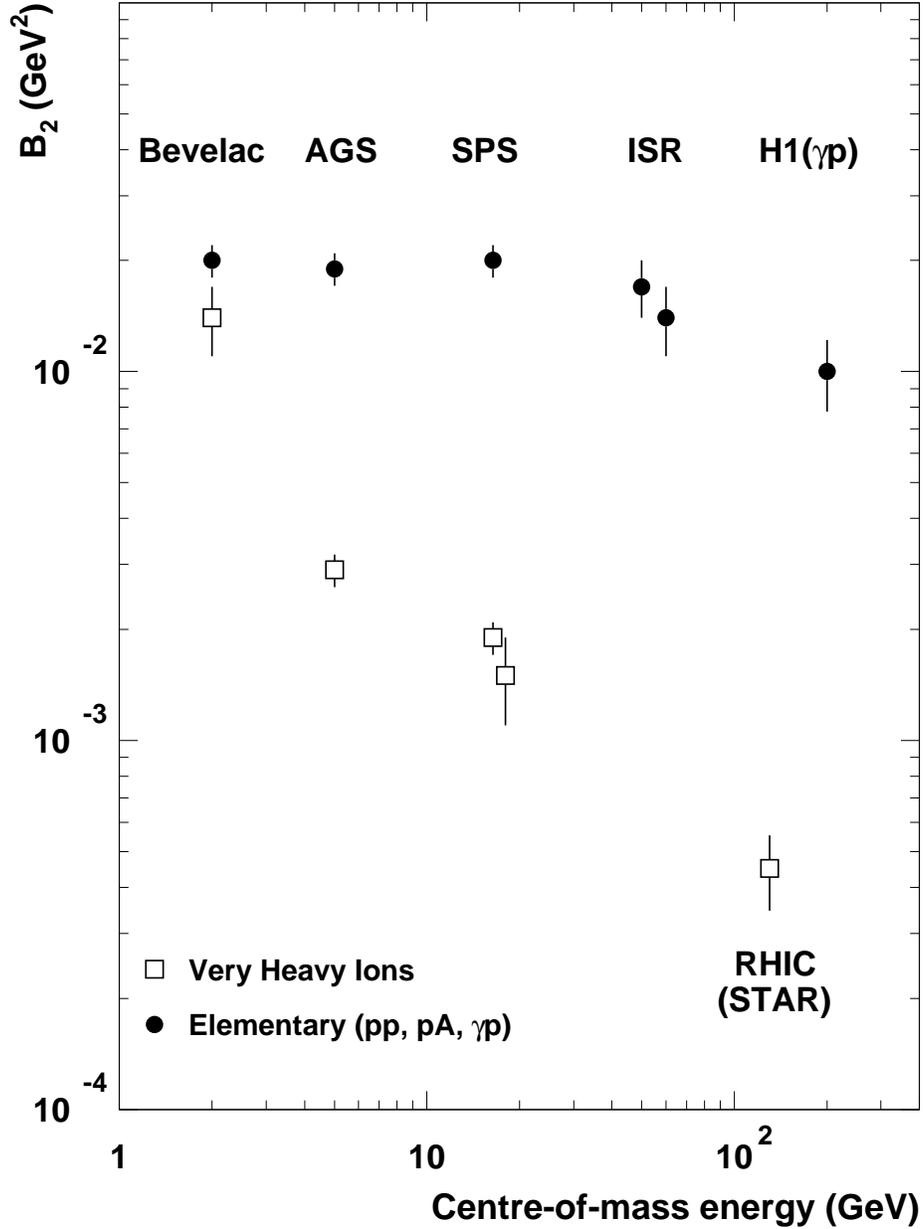,height=18cm}
\caption{The dependence of $B_2$ on centre-of-mass energy for very heavy ion 
collisions (open squares) and interactions of more elementary particles 
(closed circles). The heavy ion data are the Ne-Au data at the 
Bevelac \cite{Bev}, the E886 Au-Pt data at the AGS \cite{E886}, the Pb-Pb 
data of NA44 \cite{NA44} and NA52 \cite{NA52} at the SPS and the Au-Au data 
of STAR \cite{STAR} at RHIC. The ``elementary'' data are 
the $pA$ data of \cite{Bev,E886,NA44p}, the $pp$ 
data at the ISR \cite{ISR1,ISR2} and the photoproduction data 
presented here. In the latter the error bar represents the total uncertainty 
i.e. the sum in quadrature of the systematic and statistical uncertainties.}  
\label{B2vsE}
\end{center}
\end{figure}

Figure \ref{b2} and table \ref{Results} show the parameter $B_2$, 
computed according to equation \ref{coal1} from the data presented here. 
This quantity is calculated for the ISR data in \cite{ISR1,ISR2}, and is 
presented in figure \ref{b2} as a function of $p_T/M$. The average 
value in photoproduction is determined from the data presented here to be 
$B_2 = 0.010 \pm 0.002 \pm 0.001 \pm 0.002$, where the first uncertainty is 
statistical, the second systematic from the errors in the efficiencies 
and the third the theoretical error from the weak decay correction. 
Figure \ref{B2vsE} shows the mean value of $B_2$ in photoproduction, 
compared with the mean $B_2$ values obtained from the ISR $pp$ data 
and the data of 
a variety of other experiments as a function of centre-of-mass energy.  
The value of $B_2$ measured in photoproduction at $W_{\gamma p}=200$ GeV 
is similar in magnitude to the values   
deduced at lower centre-of-mass energies in $pp$ and $pA$ 
interactions (labelled ``elementary'' in figure \ref{B2vsE}). 
However, this value of $B_2$ is over an order of magnitude larger than 
that observed in Au-Au collisions at RHIC at a similar nucleon-nucleon 
centre-of-mass energy. Comparison of the heavy ion data with  
the data from more elementary targets shows that this discrepancy 
grows with centre-of-mass energy (see figure \ref{B2vsE}). 
To illustrate the difference between light and heavy colliding 
particles, the heavy ion data in figure \ref{B2vsE} are restricted to very 
heavy ions. They are also restricted, at centre-of-mass energy below 50 GeV, 
to measurements of inclusive deuteron and proton production to avoid 
threshold effects in anti-deuteron production. The Bevelac data, at 
which energy the $A$ dependence is weak, are the Ne-Au measurements 
of \cite{Bev}. The AGS data are the Au-Pt measurements 
of the E886 experiment \cite{E886}, the SPS data are the Pb-Pb measurements of 
NA44 \cite{NA44} and NA52 \cite{NA52} and the RHIC data are from the 
Au-Au measurements of the STAR Collaboration \cite{STAR}. 
The ``elementary'' data are the $pA$ data of \cite{Bev,E886,NA44p}, 
the $pp$ data at the ISR 
\cite{ISR1,ISR2} and the photoproduction data presented here.    

In the coalescence model for heavy ion collisions, the parameter $B_2$, 
varies inversely with both the volume of the 
fireball at thermal 
freeze-out and the absolute rate of anti-nucleon production 
when both the spatial and momentum dependence are included in the model
\cite{TS1}. The $\bar d$ to $\bar p$ ratio is observed  
in $pp$ and $\gamma p$ collisions to be close to that in Au-Au 
collisions, yet there is a much larger anti-nucleon production rate, 
and a smaller value of $B_2$ in Au-Au collisions.  These facts can be 
reconciled in the coalescence model if the size of the fireball at 
thermal freeze-out in $pp$ and $\gamma p$ collisions is much smaller than that 
in Au-Au collisions.  

\section{Conclusions}
A search for heavy charged particles is made in photoproduction at HERA and   
anti-deuterons are observed at $\langle W_{\gamma p} \rangle = 200 \gev$.  
Upper limits at the $95\%$ confidence level on the production cross sections 
for any type of positive particles heavier than tritons or negative particles 
heavier than  anti-deuterons are set at 0.19 nb in the kinematic range 
$|y| < 0.4$ and $0.2 < p_T/M < 0.7$.  The total cross section for 
anti-deuteron photoproduction is measured to be 
$2.7 \pm 0.5 \pm 0.2~ \rm {nb}$ 
in the same kinematic range and the ratio of the number of anti-deuterons 
to anti-protons is measured to be $(5.0 \pm 1.0 \pm 0.5) \cdot 10^{-4}$ in 
the range $|y| < 0.4$ and $0.3 < p_T/M < 0.7$.  
The transverse momentum dependence of the normalised invariant  
cross section for anti-deuteron production is found to be compatible with 
that measured in central $pp$ interactions at a centre-of-mass energy of 
$53 \gev$. The production rate per event of anti-deuterons in photoproduction 
is found to be over two orders of magnitude less than that observed in 
Au-Au collisions at RHIC, although the ratio of anti-deuterons to 
anti-protons is only slightly smaller. The coalescence model parameter 
$B_2$ is extracted in photoproduction and shown to be similar 
to that deduced from central high energy $pp$ data and lower energy $pA$ 
data. Averaging over the measurement range of $p_T/M$, $B_2$ is measured 
in photoproduction at $W_{\gamma p}=200$ GeV to be 
$0.010 \pm 0.002 (\textrm{stat.}) \pm 0.001 (\textrm{sys.}) 
\pm 0.002 (\textrm{theory})~\gevsq$. This value is much larger than 
that observed in Au-Au collisions at RHIC at a similar nucleon-nucleon 
centre-of-mass energy. This difference between heavy ion and elementary
particle collisions is reduced as 
the centre-of-mass energy decreases. These observations can be understood 
within the framework of the coalescence model if the interaction 
volume at thermal freeze-out in $\gamma p$ and $pp$ collisions 
is much smaller than that in Au-Au collisions at a centre-of-mass energy of 
200 GeV.

\section*{Acknowledgments}

We are grateful to the HERA machine group whose outstanding efforts have made
this experiment possible.  We thank the engineers and technicians for their
work
in constructing and now maintaining the H1 detector, our funding agencies for
financial support, the DESY technical staff for continual assistance and the
DESY directorate for support and for the hospitality which they extend to the
non-DESY members of the collaboration.



%


\clearpage

\begin{table}[htb]
\caption{ The numbers ($N_{ik}^{\gamma p, pG, eG})$ 
of photoproduction ($\gamma p$) and beam 
gas ($pG, eG$) hard selected tracks ($\log_{10} \dedx > 0.4$)     
in the four event samples (see text) 
for the different particle types. The right hand column gives the 
observed total number of identified tracks in each category after 
subtraction of the material background. The quoted 
errors are the statistical uncertainties. }
\begin{center} 
\begin{tabular}{|c|c|c|c|c|c|c|}
\hline
  & Sample & $N^{\gamma p}_{ik}$ & $N^{pG}_{ik}$ & $N^{eG}_{ik}$ && $N_{ik}$ \\
\hline
$p$ & C11 &$ 62883\pm 510$ &$1177\pm 350$&---& & $64060\pm 370$ \\
$p$ & C10 &$ 1858\pm 520$ &$7794\pm 510$&---& &$9652\pm 124$ \\
$p$ & C01 &$ 1153\pm 86$ &---&$322\pm 71$& & $1475\pm 51$ \\
$p$ & C00 &$ 25\pm 10$ &$80\pm 45$&---& & $105\pm 11$ \\
\hline
$p$ &$ N^{\gamma p,pG,eG}$ &$ 65919 \pm 733$ &$ 9051 \pm 620$ &$322 \pm 71 $ &&$ 75292 \pm 394 $\\ 
\hline
\hline
$d$ & C11 &$ 15\pm 56$ & $137\pm 41$ &--- &&$152\pm 44$ \\
$d$ & C10 &$ 0\pm 1$ & $908\pm 38$ &---&& $908\pm 38$ \\
$d$ & C01 &--- &---&$33\pm 7$ && $33\pm 7$ \\
$d$ & C00 & --- & $5\pm 2$ &---&&$5\pm2$ \\
\hline
$d$&$ N^{\gamma p,pG,eG}$ &$ 15\pm 56$ &$ 1050\pm 60$&$33\pm 7$ 
&&$ 1098\pm 60$\\ 
\hline
\hline
$t$ & C11 &$ 1\pm 9$ &$10\pm 3$&---&&$11\pm 9$ \\
$t$ & C10 & --- &$68\pm 9$&---&&$68\pm 9$ \\
$t$ & C01 & --- &---&---&&--- \\
$t$ & C00 & --- &---&---&&--- \\
\hline
$t$&$ N^{\gamma p,pG,eG}$ &$ 1\pm 9$ &$ 78\pm 10$&---&&$ 79\pm 13$\\ 
\hline
\hline
$\bar p$ & C11 &$ 61949\pm 252$ &$1\pm 9$&---& & $61950\pm 252$ \\
$\bar p$ & C10 &$ 1425\pm 71$ &$5\pm 60$&---& & $1430\pm 38$ \\
$\bar p$ & C01 &$ 1363\pm 70$ &---&$-15\pm 60$& & $1348\pm 37$ \\
$\bar p$ & C00 &$ 21\pm 5$ &---&---& & $21\pm 5$ \\
\hline
$\bar p$ &$ N^{\gamma p,pG,eG}$ &$ 64758 \pm 270 $ 
&$ 6 \pm 60$ &$ -15 \pm 60$&&$ 64749 \pm 255$\\ 
\hline
\hline
\vspace*{1mm}
$\bar d$ & C11 &$ 43 \pm 7 $ &---&---&&$43 \pm 7$ \\
$\bar d$ & C10 &$ 2 \pm 1.4 $ &---&---&&$2 \pm 1.4$ \\
$\bar d$ & C01 &--- &---&---&&--- \\
$\bar d$ & C00 &--- &---&---&&--- \\
\hline
\vspace*{1mm}
$\bar d$ &$ N^{\gamma p,pG,eG}$ &$ 45 \pm 7$ &---&---&&$ 45 \pm 7$\\ 
\hline

\end{tabular}
\end{center}
\label{Table1}

\end{table}


\begin{table}
\caption{ The observed number of events, track and event efficiencies and 
the differential 
cross section for $\bar d$ production (see section \ref{effys}).}
\begin{center} 
\vspace*{-3mm}
\hspace*{-10mm}
\begin{tabular}{|c|c|c|c|c|c|}
\hline
 $P_T/M $ & $0.25$ & $0.35$ & $0.45$ & $0.55$ & $0.65$ \\
\hline
 $N^{obs}_{\bar d}$ & 11 & 11 & 5 & 5 & 3 \\
 $\epsilon_{(\dedx)}$ & 1.0 & 1.0 & 1.0 & $0.97 \pm 0.01$ & $0.75 \pm 0.03$ \\
 $\epsilon_{\phi}$ &$0.97 \pm 0.01$ & $0.97 \pm 0.01$ & $0.97 \pm 0.01$ & $0.97 \pm 0.01$ & $0.97 \pm 0.01$ \\
$\epsilon_{cut}$ & $0.83 \pm 0.05 $ & $0.96 \pm 0.02$ & $1.0$ & $1.0$ & $1.0$ \\
 $\epsilon_{\sigma}$ & $ 0.80 \pm 0.07$ & $0.82 \pm 0.07$ & $0.85 \pm 0.07$ & $0.86\pm0.07$ & $0.87 \pm 0.07$ \\
 $\epsilon_{hit}$ & $0.67 \pm 0.04$ & $0.86 \pm 0.03$ & $0.86 \pm 0.03$ & $0.86\pm0.03$ & $0.86 \pm 0.03$ \\
 $\epsilon_{trig}$ & $0.82 \pm 0.04$ & $0.82 \pm 0.04$ & $0.82 \pm 0.04$ & $0.82\pm0.04$ & $0.82 \pm 0.04$ \\
 $\epsilon_{tag}$ & $0.46 \pm 0.02$ & $0.46 \pm 0.02$ & $0.46 \pm 0.02$ & $0.46\pm0.02$ & $0.46 \pm 0.02$ \\
 $\epsilon_{Nch}$ & $0.96 \pm 0.02$ & $0.96 \pm 0.02$ & $0.96 \pm 0.02$ & $0.96\pm0.02$ & $0.96 \pm 0.02$ \\
$\epsilon_{t0}$ & $0.95 \pm 0.02$ & $0.97 \pm 0.01$ & $0.98 \pm 0.01$ & $0.98\pm 0.01$ & $0.98 \pm 0.01$ \\
$\epsilon_{PhSp}$ & $0.75 \pm 0.06$ & $0.96 \pm 0.03$ & $1.0$ & $1.0$ & $1.0$ \\
\hline
$\frac{\rm d\sigma}{\rm dp_T/M}(nb)$  & $13.1 \pm 3.9 \pm 2.2$ & $6.6 \pm 2.0 \pm 0.8$ & $2.7 \pm 1.1 \pm 0.3$ & $2.7 \pm 1.1 \pm 0.3$ & $2.1 \pm 1.2 \pm 0.2$ \\
\hline
\end{tabular}
\end{center}
\label{dbex}
\end{table} 

\begin{table}
\begin{center}
\caption{The measured values of the invariant $\bar d$ production 
cross sections, $\bar d$ to $\bar p$ ratios and coalescence parameter,  
$B_2$. The first error is the statistical and the second error the 
systematic uncertainty. } 
\begin{tabular}{|c|c|c|c|c|}
\hline
$p_T/M$ & $\gamma \frac{\rm d^3 \sigma_{\bar d}}{\rm d^3(p/M_{\bar d})}$ (nb) &
$R_{meas}=\frac{N_{\bar d}}{N_{\bar p}} \cdot 10^4$  & $R_{corr} \cdot 10^4$ & $B_2 (\gevsq$) \\
        &    & measured value & weak decay corrected & \\
\hline 
$0.25$ & $10.5 \pm 3.2 \pm 1.7 $  & --- & --- & --- \\
\hline
$0.35$ & $3.6\pm 1.1 \pm 0.5 $ & $9.0 \pm 2.7 \pm 1.2$&$11.8 \pm 3.6\pm 1.6$ & $ 0.015 \pm 0.004 \pm 0.002 $  \\
\hline
$0.45$ & $1.2\pm 0.5 \pm 0.1 $ & $3.6 \pm 1.6 \pm 0.4$ &$4.7 \pm 2.1 \pm 0.5$ & $ 0.007 \pm 0.003 \pm 0.001 $  \\
\hline
$0.55$ & $1.0\pm 0.4 \pm 0.1 $ & $3.8 \pm 1.6 \pm 0.5$&$4.8 \pm 2.0 \pm 0.6$ & $ 0.009 \pm 0.004 \pm 0.001 $  \\
\hline
$0.65$ & $0.6\pm 0.4 \pm 0.1 $ & $3.3 \pm 1.9 \pm 0.4$&$4.2 \pm 2.4 \pm 0.5$ & $ 0.010 \pm 0.006 \pm 0.001 $  \\
\hline
\end{tabular}
\label{Results}
\end{center}
\end{table}


\end{document}